\documentstyle[12pt,amssymb]{article}
\makeatletter
\@addtoreset{equation}{section}
\makeatother

\def\dalemb#1#2{{\vbox{\hrule height .#2pt
        \hbox{\vrule width.#2pt height#1pt \kern#1pt
                \vrule width.#2pt}
        \hrule height.#2pt}}}

\def\tA{\widetilde A}

\def\0{{\sst{(0)}}}
\def\1{{\sst{(1)}}}
\def\2{{\sst{(2)}}}
\def\3{{\sst{(3)}}}
\def\4{{\sst{(4)}}}
\def\5{{\sst{(5)}}}
\def\6{{\sst{(6)}}}
\def\7{{\sst{(7)}}}
\def\8{{\sst{(8)}}}

\def\tV{\widetilde V}

\def\tA{\widetilde A}

\def\Z{\rlap{\sf Z}\mkern3mu{\sf Z}}
\def\R{\rlap{\rm I}\mkern3mu{\rm R}}
\def\G{{\cal G}}

\def\CS{{\cal S}}

\def\wtd{\widetilde}

 \def\bd{\begin{document}} \def\ed{\end{document}}
\def\ds{\documentstyle} \let\fr=\frac \let\bl=\bigl \let\br=\bigr
\let\Br=\Bigr \let\Bl=\Bigl
\let\bm=\bibitem
\let\na=\nabla
\let\pa=\partial \let\ov=\overline
\newcommand{\be}{\begin{equation}}
\newcommand{\ee}{\end{equation}}
\def\ba{\begin{array}}
\def\ea{\end{array}}
\def\ft#1#2{{\textstyle{{\scriptstyle #1}\over {\scriptstyle #2}}}}
\def\fft#1#2{{#1 \over #2}}
\def\del{\partial}
\def\sst#1{{\scriptscriptstyle #1}}
\def\oneone{\rlap 1\mkern4mu{\rm l}}
\def\ie{{\it i.e.\ }}
\def\via{{\it via}}
\def\semi{{\ltimes}}
\def\v{{\cal V}}
\def\str{{\rm str}}

\newcommand{\ho}[1]{$\, ^{#1}$}
\newcommand{\hoch}[1]{$\, ^{#1}$}
\newcommand{\bea}{\begin{eqnarray}}
\newcommand{\eea}{\end{eqnarray}}
\newcommand{\ra}{\rightarrow}
\newcommand{\lra}{\longrightarrow}
\newcommand{\Lra}{\Leftrightarrow}
\newcommand{\ap}{\alpha^\prime}
\newcommand{\bp}{\tilde \beta^\prime}
\newcommand{\tr}{{\rm tr} }
\newcommand{\Tr}{{\rm Tr} }
\newcommand{\NP}{Nucl. Phys. }
\newcommand{\ens}{\it Laboratoire de Physique Th\'eorique de l'\'Ecole
Normale Sup\'erieure\hoch{3}\\
24 Rue Lhomond - 75231 Paris CEDEX 05}

\newcommand{\auth}{ B. Julia}
\begin{document}
\begin{flushright}
\hfill{LPTENS-00/02  }\\
\hfill{hep-th/0002nnn}\\
\hfill{Feb. 4, 2000}\\
\end{flushright}
\begin{center}
{ \large {\bf Superdualities: }}

{\large {\bf Below and beyond U-duality \hoch{1,2}}}

\vspace{5pt}
\auth

\vspace{5pt}

\ens

\vspace{5pt}

\underline{ABSTRACT}
\end{center}
Hidden symmetries are the backbone of Integrable two-dimensional theories. They
provide classical solutions of higher dimensional models as well, they  
seem to survive partially quantisation and 
their discrete remnants in M-theory called U-dualities, 
would provide a way to control infinities and nonperturbative effects in 
Supergravities and String theories. Starting from Einstein gravity
we discuss the building blocks of these large groups of internal symmetries,
and embed them in superalgebras of dynamical 
symmetries. The classical field equations for all bosonic matter fields of 
all toroidally compactified supergravities are invariant under such 
``superdualities''. Possible extensions are briefly discussed.

{\vfill\leftline{} \vfill
\footnoterule
{\footnotesize \hoch{1} Based on talks delivered at the International 
Seminar "Supersymmetries and Quantum
 Symmetries", JINR Dubna (27-31 July, 1999), XXII-nd TRIANGLE MEETING Utrecht 
(26-28 May 1999), IHES workshop on D-Branes, 
Vector Bundles and Bound States (July 1999) and Tel-Aviv Workshop on Recent 
Developments in String Theories, Conformal Field Theories and Integrable Models 
(7-11 January 2000).\vskip -12pt}\vskip14 pt
{\footnotesize \hoch{2} Supported by EC under TMR
contracts ERBFMRX-CT96-0045 and FMRX-CT96-0012 \vskip -12pt} \vskip 14pt
{\footnotesize
        \hoch{3} UMR 8549 du
CNRS et de l'{\'E}cole Normale Sup{\'e}rieure. 
\vskip -12pt}}

\pagebreak
\setcounter{page}{1}

\section{Introduction}

Since the discovery of discrete U-dualities, of their role in the control of
the divergences of string theories \cite{HTo,To,Wi} and of the duality
between Large N super-Yang-Mills theory  and AdS compactification of
eleven dimensional supergravity \cite{Ma} the need for a better
conceptual understanding of the latter has become rather    urgent.
The superspace approach is notoriously hard but component approaches
are
rather cumbersome, this is unsatisfactory as more miracles are  being
discovered \cite{CJLP2}.
These dualities are   important also  to
clarify the constraints on allowed counterterms
and nonperturbative effects in Supergravities see \cite{GV, DJST} and
references therein.

In section 2 we shall review the relation between the background
spacetime geometry  and 
the duality symmetries. This is a long and pedestrian approach towards the 
background independent stucture of M-theory. 
It turns out that all massless bosonic matter fields of the toroidally 
 compactified theory obey one rather simple universal self-duality classical
equation of motion. The self-duality involution is the product of
Hodge duality on all forms (the bosonic matter fields) by an internal twist
which in particular 
 compensates for the noninvolutive character of Hodge duality in some 
dimensions of spacetime. More conceptually the twist exchanges generators and 
their conjugates in a doubled superalgebra  (still in the bosonic sector)
that captures all the nonlinearities. In section 3 we shall recall previous 
instances of this self-duality in even dimensions $D=2k$ for the k-forms,
and proceed to generalise it to all forms and all dimensions following 
\cite{CJLP2}. The  superalgebra of dualities contains as subalgebra the 
symmetry of a supertorus with one fermionic dimension on top of the 
compactification torus one is assuming. In the next section we shall discuss 
the importance of 
deformation theory  for the construction of SUGRAS and discuss the deformation 
that leads to M-self-duality. Finally we shall comment on possible extensions.

\section{Ehlers' and other symmetry enhancements}
\subsection{From stationary gravity to Kaehler moduli}

Let us start our discussion with the realistic (low energy) model of four 
dimensional gravity. It is of course invariant under diffeomorphisms and if 
needed under local Lorentz transformations. This possibility reflects the fact 
that gauge restoration may increase the symmetry, it is well known that 
gauge unfixing often makes other symmetries manifest but 
 we shall see that changes of our choice of fundamental 
fields also modify the faithful symmetry group. If one considers the space of 
solutions admitting one non-lightlike Killing vector, it turns out that 
diffeomorphisms of the cyclic coordinate become gauge transformations of the 
abelian connection defined by the orthogonal hyperplanes to the Killing orbits:
local domains become fibered by the orbits and inherit a principal connection.
Actually the global (or rigid) rescalings of the cyclic coordinate imply 
also a scaling symmetry $\R$. 
(More generally dimensional reduction on $T^k$ implies an internal symmetry
$GL(k,\R)$).
 If one however dualises the vector potential in the remaining 
three dimensions to a scalar field $B$ defined up to a gauge freedom, namely 
the addition of a constant, then the two degrees of freedom of the graviton 
conspire to parameterise the Poincar\' e upper half-plane and the abelian gauge 
invariance of the connection disappears to leave room for a rigid $SL(2,\R)$ 
of internal symmetries. This group I called the
 Ehlers group although the original 
name was given to its maximal compact subgroup $SO(2)$. The latter is the only
true surprise as the other two generators of $SL(2,\R)$ are the rescalings and 
the constant shifts of the dual $B$ field. In fact its action
 is highly nontrivial
(if somewhat singular) as it for example transforms an asymptotically flat 
Schwarzschild black hole into a Taub-NUT spacetime. These transformations act 
nonlocally on the original four-dimensional fields.  

More instances of this miracle occur in other theories, Einstein-Maxwell 
theory 
reduced from 4 to 3 dimensions has the structure of the nonlinear sigma
model $SU(2,1)/S(U(2)\times U(1))$ with four freedoms whereas one rescaling and 
two shifts are predicted \cite{kin}. 
Similarly compactification of pure gravity from d dimensions to 3 leads to an 
internal symmetry $SL(d-2,\R)$ whereas one expects $GL(d-3,\R)$. 
Supergravities representing the effective 
low energy theories of type I or heterotic strings (compactified on $T^6$)
have been conjectured in 1990 to exhibit also 
a non-perturbative (in the string-string coupling constant $g_{string}$)
 so-called S-duality, namely $SL(2,\Z)$ inside the Lie group $SL(2,\R)$. The 
latter has been  known in 
the classical 
supergravity context since the construction of the $N=4$ SUGRA in 4 
dimensions and 
the analysis by Chamseddine of the reduction of type I 10d SUGRA on $T^6$.
In this last case a form  (here a 2-form) can again
 be dualised to a scalar (axion or 
Kalb-Ramond) field to parameterise the Poincar\' e upper half-plane. 
We refer the reader to some reviews 
for more references: see \cite{J85,S93,L98} for instance. 
 Let us recall also that in \cite{b94} the same
S-duality in four dimensions is conjugated to a subgroup of
Ehlers' type namely the  $SL(2,\Z)$ associated to
a seventh Killing vector 
by the T-duality corresponding to its direction. 
T-duality is a discrete symmetry of string theories associated to an internal
isometry. It does act on the string interactions but in perturbative way and 
it acts nonperturbatively on the geometry by inverting the radius of the 
compactification circle in string units.
 
Typically the Lie group
(over the real numbers) is a symmetry of the low energy effective SUGRA type
action and the discrete arithmetic group over $\Z$ is its quantum remnant and 
it is believed to be the (gauge) symmetry of the full theory.
More generally S-dualities are those dualities that exchange weak and strong 
string-string couplings. The discreteness of  S-dualities is the 
non-perturbative 
 result of string-string interactions: for instance in the IIB theory
non-holomorphic
S-modular Eisenstein series appear as coefficients of the
 (10 dimensional) gravitational coupling expansion \cite{G99} where the string 
length appears as a cut-off.
It has been shown 
 that S-duality  of the heterotic string on $T^6$ is 
``string-string dual'' to T-duality of type IIA on $K3\times T^2$, 
in fact this 
correspondence between string theories exchanges perturbative and 
``non-perturbative'' dualities. 

Yet another enhancement to $SL(2,\R)$ occurs by $T^2$ compactification of
string theories, the complex structure modulus of the torus is as expected
a coordinate on the Poincar\' e upper half-plane, it corresponds to a 
``geometrical''  $SL(2,\R)$. At the string level, \ie with all massive
 states included,
the continuous geometric 
Lie group acts on the moduli  and only its discrete subgroup
$SL(2,\Z)$ is a quantum gauge (but perturbative) symmetry. But  
the background 2-form flux or integral over the torus  and its volume
combine to 
form its Kaehler modulus. The latter parameterises another $SL(2,\R)/SO(2)$
\cite{D88}. For instance  in the type I case reduced to 8 dimensions on top of 
the geometric 
 $GL(2,\R)$ invariance of the tangent space to the compactification
 torus which is relevant in the zero mass sector there
appears another $SL(2,\R)$, in the type IIA case the same group emerges. From 
the M-theory point of view the flux of 
the three form and the volume of $T^3$ ($SL(3,\R)$ scalars) make a similar 
complex valued modulus and in the type 
IIB case the 2-form fluxes enhance the $SL(2,\R)$ already present in 10 
dimensions to $SL(3,\R)$ commuting with the 
diffeomorphism invariance induced $SL(2,\R)$.
In these examples double T-duality of the torus $T^2$
is the nontrivial generator of the extra $SL(2,\Z)$ that commutes with the
obvious (geometric) $SL(2,\R)$.

 Let us note that
the T-duality relating type IIA and type IIB exchanges 2 rather different 
dimensions, 
not only are  the lengths of the dual circles inversely proportional 
but in 8 dimensions the $SL(3,\R)$ coming from unimodular 
diffeomorphisms of the M-theory compactified on $T^3$ commute
with the $SL(2,\R)$ coming from unimodular 
diffeomorphisms of type IIB compactified on $T^2$, although one dualises only 
one direction. this means that our classical approximation of spacetime
and even its number of dimensions are model dependent, an important issue is to 
determine the domain of validity of each of these ``complementary'' (in the 
sense of Bohr) classical approximations, in that connection see \cite{J98}. 

For completeness let us recall that the $SL(2,\R)$ symmetry of IIB SUGRA in 
10 dimensions  may be traced back to its 12 dimensional origin, 
namely F-theory 
compactified on a torus $T^2$ with 
frozen Kaehler modulus, in particular this means that in 12 dimensions if there
is diffeomorphism invariance it is only for the volume preserving subgroup, I 
have emphasized this point in my study of the group disintegration of $E_8$
in for instance \cite{J98}.  
We note here that the group of unimodular diffeomorphisms is precisely
 the invariance group of
the action of isentropic perfect fluids (compressible or not)
 expressed in Lagrangian coordinates.  
In fact if we fix the volume of $T^2$  the 
large radius limit of one of its directions corresponds to a
 small radius limit for 
the other one and hence effectively 11 dimensions not 12. Another peculiar 
compactification along a
torus $T^2$ of null radii produces IIB superstring theory 
from M theory, this was analysed by Aspinwall and leads to the 
interpretation of the  $SL(2,\Z)$ invariance  of IIB string theory in 10
dimensions as the geometric invariance of the torus.
One of the null radii is actually infinite on the  IIB side after T-duality, 
 so the limit corresponds to a ten dimensional theory.
 
\subsection{More ``accidents''}
Let us now review quickly the build-up of the large U-duality symmetry groups 
when one increases the dimension of the compactification torus. The geometrical
 symmetries grow regularly as expected but beyond the Ehlers type accidents 
listed above even more dramatic symmetries drop out of the low energy 
analysis. For instance the M-theory (11d supergravity) symmetries are the 
Lie goups of E type suitably defined for low rank or rank 9 and maybe rank
10(?). After reduction on $T^n$ one obtains the split real form of 
$E_n$ the algebra is non simple for $1\leq n \leq 3$ and beyond that it 
becomes simple by glueing of the Ehlers type factor to the geometric symmetry.
Specifically $A_2\times A_1=E_3$ becomes $A_4=E_4$. Other glueings occur 
for type I SUGRA on $T^3$ where $D_2$ becomes $D_3$ and on $T_7$ where
$D_6\times A_1$ becomes $D_8$. In that family the algebras of type D appear
also in their split form.  In the string context they also appear \cite{HJ}.

It is a 
classical result  that $SO(n,n+k)$ symmetries occur on $T^n$ if one starts from
type I SUGRA coupled to k vector multiplets in 10 dimensions. The groups
$SO(n,n+16)$ correspond to heterotic strings where 16 is the rank of the 
internal gauge group.

In all cases the symmetry extends to the affine (Kac-Moody) Lie algebras 
corresponding to the three dimensional theory upon further reduction to 
two dimensions. At the classical level, the action has been
 successively found to be non symplectic (see for instance the discussion and 
references in \cite{J98}) and Lie-Poisson \cite{S1,S2} leading at
the quantum level to the  
proposal to use a quantum group, the stringy version remains
unknown. The interplay between arithmetic, affine and quantum groups deserves 
more study. 

Conversely one might ask oneself whether a curved space sigma model in three
dimensions, namely (topological) gravity coupled to a symmetric space sigma 
model (notwithstanding any supersymmetry) is actually the result of a 
toroidal dimensional reduction. This has been studied extensively in the last 
century \cite{J99}. The result is strikingly simple: the rule of group 
disintegration (also called oxidation) \cite{J81} requires that the affine
 Dynkin diagram of the 3d group contains a geometric $SL(D-3)\times \R$ 
ending at the affine root for 
an ancestor theory to exist in dimension D. This had been used to predict a
new SUGRA in 6 dimensions that was actually constructed recently \cite{K97}.
If one considers the scalar sectors of the $E_n$ family, but one puts them
 now in 3 curved dimensions (we stick to  the split forms, this ruins most 
supersymmetries) then one obtains a surprising magic 
triangle of higher dimensional ancestors. 
The geometry of that triangle still escapes mathematicians
\cite{J00}. 
 
\section{Self-duality equations}
\subsection{k-forms in 2k dimensions}
It is well known that self-duality requires $d=4k+2$ on a spacetime of
Minkowskian signature, or $d=4k$ for Riemannian spaces. The discovery of
instantons in 1975 has launched a search for exact classical solutions of 
nonlinear equations including Einstein and Yang-Mills equations. It followed
the discovery of regular magnetic monopole (and dyon) solitons and was 
concomitant with that of
 the corresponding self-dual solutions in the so-called 
Prasad-Sommerfield-Bogomolny (BPS) limit, (actually the pseudoparticle paper 
appeared precisely between the PS and the B papers and the latter emphasized 
the stability aspect). More recently  the BPS conditions 
gained importance because of their realisation as conditions for unbroken 
supersymmetry. 
As a toy example, the scalar  wave equation in two dimensions 
admits self-dual solutions, the left and right moving modes. The 
(i)-self-duality equation on a Riemann surface 
$ df=i*df $
implies harmonicity. 

Conversely the Cauchy-Riemann equations are related to the real solutions
 of the harmonic equation, they can be seen as 
a twisted self-duality equation for a pair of functions ($Ref=a$ and $Imf=b$):
$ da=*db$
a and b are conjugate harmonic functions, this fact has been used in \cite{J96}
to render the action of the $SL(2,\R)$ subgroup of the conformal group 
manifest: $a$ and $ b$ are coordinates on the Poincare half-plane again.
Let us introduce the doublet $C=(a,b)$, the above equation can be rewritten
\be
dC=*dC S .
\label{F1}
\ee
Note that we chose euclidean signature and preserved 
the reality of $C$ nevertheless, the twisting matrix $S:=i\tau_2$ has square 
$-1$ precisely to compensate the annoying $**=-1$. To summarize the procedure,
we replaced one second order equation for one unknown, here the harmonic 
equation for $a$, by two first order
equations for two unknowns which are equivalent to the original problem.
The first order system is then shown to possess a rigid symmetry, the duality 
$S$ that acts locally on the pair of field strengths
but nonlocally on the original 
function a. This despairingly simple example is actually the prototype of
our final result.
  Strictly speaking there are topological and gauge subtleties because the data
of $C$ is not equivalent to that of $a$ there is an integration constant to
be handled by a ``normalisation'' condition. 

The Geroch group action on 
Einstein plane waves is realised on an infinite set of potentials related 
also by duality equations called Baecklund transformations and usually combined
into a linear system depending on one parameter. The consistency conditions
are the original equations and the corresponding equations for the dual 
fields. But let us stay in two effective
(``active'') dimensions and consider the principal sigma model for a 
semi-simple group $G$. We shall identify the  Lie algebra $\cal G$
and its dual 
by the Killing form (which appears in the action).
The equations read:
\bea 
A:=dg g^{-1}=A_cT^c \\
dA-A\wedge A\equiv 0 \\
d*A-A \wedge *A +*A \wedge A = 0
\label{sigma}
\eea
In order to exhibit the symmetry between the Bianchi identity and the equation 
of motion one is led to define dual generators $\tilde T^c= S(T^c)$ and to 
impose $(S*)^2 ={\rm Id.}$ by defining suitably the action of the linear
``involution'' S on the dual generators.
The equations now read 
\bea
dC-C\wedge C\equiv 0 \\
C=S(*C)
\label{F2}
\eea
provided $T^c$ and $\tilde T_c$ form the Lie algebra $\G \semi \G^*$. Note
that the Killing form is only used for the definition of $S$.
We see the second example of a universal twisted self-duality equation
that encodes the full original second order system. 

Our next example comes from abelian vector fields in maximal supergravity in 
4 dimensions. This has been analysed first
with gauge fixed coset in $N\leq 4$ SUGRA 
\cite{Z,CF} and then extended to include the maximal SUGRA  with the general 
gauge invariant structure under the maximal compact subgroup 
\cite{CJ,J98}. Shortly thereafter this technology has been transferred to
statistical mechanics.
 The duality symmetry $E_7$ can only act on the doubled set
of the fundamental 28 potentials plus their 28 dual potentials that together 
form the 56 representation  of $E_7$:  the abelian one forms ${\cal C}$.
They obey again an analogous system of equations:
\bea
d{\cal C}\equiv 0\\
V{\cal C} =*S V{\cal C} 
\label{F3}
\eea   
where $S^2=-{\rm Id.}$ and $V$ is the scalar matrix representing
$E_7$ in the fundamental representation 56.
This structure extends to higher dimension $d=2k$ for k-forms which now
fall into representations of  the 
groups $E_5=D_5$ and $ E_3$ \cite{CJLP1}.

\subsection{11d SUGRA has a twelfth fermionic dimension}

Dualisations of all forms is possible in the toroidal compactifications of
11d SUGRA \cite{cjs,CJLP1,CJLP2} or at least the equations of motion for the
doubled set take a nice form. The dualisability of the 3 form was discovered 
in 10 dimensions long ago, but  
for the eleven dimensional theory although one can double the set of fields 
one cannot dualise the three  form, see \cite{GN,L} and references
 therein. In fact 
independently of our group doubled Lagrangians were proposed \cite{S} but they
do 
not exhibit the nonabelian structure that generalises the semi-direct product 
algebra of sigma models presented above.

The bosonic action we shall consider in 11d reads:
\be
{\cal L}_{11} = 
\kappa^{-2} R\, {*\oneone} - \ft12 {*F_\4} \wedge F_\4 -\kappa \ft16
F_\4\wedge F_\4 \wedge A_\3\ ,\label{d11lag}
\ee
the matter equations can be rewritten in our favorite form  
by introducing the dual 6-form $\tA_\6$:
\be
     {*\G} = \CS \, \G\ \label{geq}
\ee
\be
d\G - \G\wedge \G \equiv 0 \label{cmeq}
\ee
provided $\G=d\v\, \v^{-1}$ and
\be
\v = e^{A_\3\, V}\, e^{\tA_\6\, \tV}\ .\label{d11coset}
\ee
In these expressions the forms should be expanded and treated as odd 
Grassmann parameters if of odd degree, then the full content of the nonlinear
Chern-Simons term (dictated by supersymmetry and proportional, with a 
specific coefficient, to the gravitational coupling constant $\kappa$) 
follows from the Clifford-type superalgebra structure:
\be
\{V,V\}= -\kappa \tV\ ,\qquad {[}V,\tV {]}=0\ ,\qquad {[} \tV, \tV {]} =0\ .
\label{d11com}
\ee
Equation \ref{d11coset} is a generalisation of ordinary sigma models, the 
rigid symmetry has become a gauge symmetry in the generalised sense of
\cite{bj80}
namely:
\be
\v' = \v\,e^{\Lambda_\3\, V}\, e^{\wtd\Lambda_\6\, \tV}\ .
\label{cosetgauge}
\ee
where $\Lambda_\3$ and $\wtd \Lambda_\6$ are closed
forms. The reader may wonder what has been gained by the replacement of a
p-form by a closed (p+1) form, it turns out that the (p+1)th de 
Rham cohomology leads to conserved generalised charges 
\cite{llps} and the superalgebra of gauge symmetries implies
 nonlinear relations of the type
\be
t_k t_l = t_{k+l}
\label{qc}
\ee
where the $t_k$ are the renormalised tensions of the $(k-1)$
branes (fundamental ones or  their duals) $t_k:=T_{k-1}/2\pi$. In the present 
situation one finds
$(t_3)^2=t_6$.
One
recovers one of the relations  found in  \cite{DLM}. More generally the
known relations between tensions are of several types: they are 
 either Dirac-Nepomechie-Teitelboim (DNT)
quantisation conditions or they express global well-definiteness of the action 
classically or absence of anomalies, finally they may also be obtained from 
T-duality as shown first by Polchinski. We would like to stress that by the 
present classical considerations and the ``single'' valuedness of the action
\cite{DLM} one recovers the DNT quantisation condition. In general eqs.
\ref{qc} are quite powerful.  

What is rewarding is that after toroidal compactifications all the equations of motion of the bosonic matter fields of maximal SUGRA's 
are reproduced (some of them really derived for the first time ab initio)
 and encoded by a relatively simple superalgebra. Note 
that some degrees of freedom of the graviton  become progressively matter
fields, eventually all of them when the theory has been reduced to 3 dimensions.
All the matter field equations do follow our universal pattern, namely
\ref{geq} and \ref{cmeq}.
The necessary superalgebra is a deformation described compactly in \cite{CJLP2}
of the following finite dimensional superalgebra:
\bea
\G:=  {\cal A}\semi {\cal A}^* \\
{\cal A}:=sl_+(n|1) \semi (\wedge w)^3
\label{F5}
\eea
where n is the dimension of the torus, the twelfth fermionic dimension 
appears in the classical superalgebra $sl(n|1)$,  $w$ is its fundamental 
representation which decomposes as $w=v+1$ (with $v$
 an n-component vector) as representation 
of $sl(n)$. Only the Borel (triangular) subsuperalgebra appeared yet 
(this is the meaning of the $+$ index), and 
its semidirect product is with the (super)antisymmetric tensor of third 
order. The latter unifies the three form and its descendent 2-forms etc.
in a single representation. The deformation we alluded to reflects the 
Clifford structure above and is proportional to $\kappa$. It is my conjecture 
that we shall find an even larger (``simple'') structure whose solvable
part will be $\G$.

As promised we have encoded all equations for the bosons (but for the graviton)
and extended the U-duality symmetry. Strictly speaking we have only recovered
the Borel subgroup of $E_n$ as follows. The superalgebra involved is 
$\Z$-graded (with nonpositive degrees) and its coefficients are forms  
whose degrees compensate those of the 
generators. In degree zero the coefficients are scalar fields and their
dual (D-2)-forms are coefficients of the generators of degree -(D-2) 
(note these are odd in odd spacetime dimensions). The degree  zero sector of 
the above superalgebra is $b_0=sl_+(n) \semi (\wedge v)^3 $, this gives
 precisely the Lie algebra of 
the scalar Borel manifold $B$ (as Iwasawa told us: for a noncompact symmetric 
space $B \approx K\backslash G$), see \cite{CJ,lpsweyl} and references therein.

\section{Towards M-theory}
\subsection{Low energy theory}
Clearly the other massless fields appropriate to the low energy approximation
should be included, firstly the graviton and then all the fermionic partners.
As far as the metric is concerned we would like to give two reasons for hope.
The first one is an old result \cite{BB} on a BPS type condition for a fourth 
order gravity theory which leads to the Einstein spaces vacuum equations. These 
equations are of second order and allow for an arbitrary cosmological constant
in four dimensions. The (twisted) self-duality equation reads again:
\be
R=*S R
\label{F6}
\ee 
where $R$ is the Lorentz algebra valued curvature 2-form (torsion is set equal 
to zero) and $S$ is the Lorentz ``Hodge'' dualisation. It is not exactly what 
we are looking for but in four dimensions and up to the cosmological constant
problem it comes close! 

Another encouraging sign is the fact that the deformation 
proportional to the gravitational coupling constant is the only mysterious 
feature in the matter sector, the undeformed superalgebras listed above are 
quite natural. The fact that the semidirect product ${\cal A}\semi {\cal A}^*$
occurs everywhere is quite reminiscent of the orbit method in group theory with 
the cotangent bundle to the group as the basic symplectic object. At the 
quantum level it could be related to some ``Quantum doubles''. 
Another hint along the same line is what we called the ``Jade rule'',
namely the property that if the commutator of two generators $ V$ and $V'$ is
equal to $V''$ the  commutator of $V$ and $S(V')$ is up to a sign $S(V'')$, 
this is clearly true for semidirect products of the type 
${\cal A}\semi {\cal A}^*$ but it is significant that 
the Jade rule is preserved by deformation!
 Furthermore if we could characterise our deformation we could envisage to 
include the graviton which in fact is also deformed from a linear (free) spin 
2 field to the nonlinear
metric by the requirement of (super)diffeomorphism invariance. We shall return 
to this observation in the next section. 

Let us now return to the fermionic fields. They quite generally transform 
under the local gauge groups ($K$ the maximal compact U-duality subgroup or
the Lorentz group $L$ for the spacetime symmetries) and are inert 
under U-dualities once the corresponding gauge invariances ($K\times L$) have
been restored. One obstacle to progress has been the absence of appropriate 
simple finite dimensional or even affine  superalgebra. In three dimensional 
maximal SUGRA the Borel subgroup of $E_8$ appears as the  degree zero part of
our symmetry superalgebra, the rest is the odd dual but as the superalgebra is 
far from simple, it has only nonpositive degrees, it does not appear in the 
tables of simple ones. This remark has consequences for our program of 
restoration of a larger supergroup beyond the present Borel part:
it is unlikely to be finite dimensional if simple.

Finally one should include the extended objects and their massive excitations,
this should reduce the symmetry to some kind of arithmetic subgroup.

\subsection{The Noether method and GDA's}

The cohomology of infinite Lie algebras is at the heart of the so-called Noether
method of construction of gauge theories like SUGRAS, it is in fact the only 
method available to derive 11d supergravity. In \cite{cjs} we resummed the 
gravitational infinite series leading to Einstein's action by using our 
knowledge of differential geometry. Then the local supersymmetry invariance 
was implemented order by order in $\kappa$ and in the absence of scalar fields 
this terminates after a finite number of steps. The result is the
trilinear term in the three form if one starts from the flat space linearised 
theory. Very much like in Yang-Mills theory one starts with abelian gauge 
invariances and a rigid non-abelian symmetry and one deforms the gauge 
invariance and its invariant action preserving the rigid symmetry. What we 
have
found here is similar, indeed the superalgebra of dualities is also deformed
into another more nonlinear one and the deformation parameter is again 
the gravitational coupling constant. The equations of motion have been
shown to be invariant.

Graded Differential Algebras (GDA's)
 have been studied in recent times in Differential 
Topology \cite{SU}, but with a restriction to simply connected manifolds, 
they have been also considered in SUGRA theories \cite{AF}
but with the same restriction (freeness to be lifted for sure) 
and finally in the analysis of BRS 
cohomology. Here the nonlinearities are different from Yang-Mills
 non linearities
instead of expressions like $F=dA-A\wedge A $ one encounters
\be
F=dA - dA'\wedge A''...
\label{DGA}
\ee 
The (super)group is obtained by exponentiating
(degree zero) combinations of degree $n$ ($n\geq 0$ for the time being) 
differential forms and degree $-n$ generators of a $\Z$-graded superalgebra.
Nonlinearities of the type \ref{DGA} have been encountered before also in the 
coupling of type I SUGRA to  abelian gauge fields, they were once more
 dictated by 
the requirement of local supersymmetry. In the nonabelian case
the Chapline-Manton coupling in a 
sense combines the present nonlinearities and those of Yang-Mills theory.   

\section{Conclusion}
The selfdual superalgebras have other applications under study, they reproduce,
simplify and extend many known results but small Mysteries have converged to a 
big one.
Let us add that the Green-Schwarz term that cancels the perturbative anomalies
of naive SUGRA is also of a (mixed) Chern-Simons nature. In \cite{HW}
the modified Green-Schwarz mechanism involves distributions defined by the 
boundary cycles but it should fit into this circle of ideas. If so   the 
relation Horava and Witten
 found between the Yang-Mills and the gravitational coupling 
constants $\lambda_{YM}^6 \propto \kappa^4$ strongly suggests a deformation 
analysis of their anomaly considerations but in terms of a parameter $\mu$
such that $\lambda \propto \mu^2$ and $\kappa \propto \mu^3$, this suggests a 
role for triality in the gravitational sector at least in 10/11 dimensions.

\section*{Acknowledgments}
I am grateful to my collaborators for sharing insights and stimulating 
discussions, and I would like to dedicate this work to the memory of Viktor 
I. Ogievetsky who shared our
interest for Group theory in Physics and for whom 
Truth stood above  careers. 
I also benefited from conversations with A. Hanany, P. Henry, V. Kac, 
H. Nicolai, B. Pioline, A. Schwarz and A. Smilga.

\end{document}